\documentclass[journal]{vgtc}                

\ifpdf
  \pdfoutput=1\relax                   
  \usepackage{graphicx}                
  \DeclareGraphicsExtensions{.pdf,.png,.jpg,.jpeg} 
\else
  \usepackage{graphicx}                
  \DeclareGraphicsExtensions{.eps}     
\fi%

\graphicspath{{figures/}{pictures/}{images/}{./}} 

\usepackage{microtype}                 
\PassOptionsToPackage{warn}{textcomp}  
\usepackage{textcomp}                  
\usepackage{mathptmx}                  
\usepackage{times}                     
\usepackage{fontawesome}
 
\usepackage{balance}  
\usepackage{cite}                      
\usepackage{tabu}                      
\usepackage{booktabs}
\usepackage{xcolor,soul}
\usepackage[caption=false]{subfig}

\usepackage{booktabs} 

\usepackage{multirow}
\usepackage{xspace}
\usepackage{colortbl}
\usepackage{bm}

\newcommand{\inlinebold}[1]{\noindent\textbf{\textit{#1.}}}
\newcommand{\strategy}[1]{\textsc{#1}\xspace}
\newcommand{\command}[1]{$\mathbf{O_{#1}}$\xspace}
\newcommand{\commandHeader}[1]{$\boldsymbol{O_{#1}}$\xspace}

\newcommand{\ourParskip}{\vspace{0.5\baselineskip}}

\usepackage{enumitem}

\usepackage{microtype} 
\usepackage{mathptmx}
\usepackage{graphicx}
\usepackage{times}
\usepackage{enumitem}
\usepackage{float}

\definecolor{resize}{HTML}{3589EC}
\definecolor{reposition}{HTML}{37C61D}
\definecolor{recolor}{HTML}{D85727}

\usepackage{tabu}

\onlineid{0}

\vgtccategory{Research}
\vgtcpapertype{please specify}

\title{Investigating Direct Manipulation of Graphical Encodings as a Method for User Interaction
}

\author{Bahador Saket, Samuel Huron, Charles Perin, and Alex Endert}
\authorfooter{
 \item Bahador Saket is with Georgia Tech. E-mail: 
 saket@gatech.edu
\item
Samuel Huron is with university Paris Saclay. E-mail: samuel.huron@telecom-paristech.fr. \item
Charles Perin is with University of Victoria. E-mail: cperin@uvic.ca.\item
Alex Endert is with Georgia Tech. E-mail: 
endert@gatech.edu
}

\shortauthortitle{Biv \MakeLowercase{\textit{et al.}}: Global Illumination for Fun and Profit}

\abstract{We investigate direct manipulation of graphical encodings as a method for interacting with visualizations. There is an increasing interest in developing visualization tools that enable users to perform operations by directly manipulating graphical encodings rather than external widgets such as checkboxes and sliders. Designers of such tools must decide which direct manipulation operations should be supported, and identify how each operation can be invoked. However, we lack empirical guidelines for how people convey their intended operations using direct manipulation of graphical encodings. We address this issue by conducting a qualitative study that examines how participants perform 15 operations using direct manipulation of standard graphical encodings. From this study, we 1) identify a list of strategies people employ to perform each operation, 2) observe commonalities in strategies across operations, and 3) derive implications to help designers leverage direct manipulation of graphical encoding as a method for user interaction. 
} 

\keywords{Direct Manipulation, Data Visualization}

\CCScatlist{ 
 \CCScat{K.6.1}{Management of Computing and Information Systems}%
{Project and People Management}{Life Cycle};
 \CCScat{K.7.m}{The Computing Profession}{Miscellaneous}{Ethics}
}

\vgtcinsertpkg

\begin{document}


\firstsection{Introduction}


\maketitle
The visualization community has advocated for the need to design more natural and fluid interactions for data visualization tools~\cite{elmqvist2011fluid}. A recent line of research investigates how to enable users to convey their intended operations by direct manipulation~\cite{shneiderman:19931} of the graphical encodings used to represent the data (e.g.,~\cite{saketVbD, chevalier:2012:histomages, kondo2014dimpvis, Perin:table, Perin:2014, Kwon:2011, riche2010understanding, Saket_VbDEval, Wenskovitch2018, desJardins:2007, kim:2016:interaxis, endert2012semantic, torres:2003:visual,Siirtola_Matrix,Baudel:2006}). 
For example, DimpVis~\cite{kondo2014dimpvis} allows users to directly interact with the length, angle and position of the visual representations, as a means for temporal navigation. With DimpVis, users can drag a bar in a bar chart vertically to adjust its height. This makes it possible to explore the values for the bar over time, and to search for times at which the bar had a particular value.
The appeal of direct manipulation can be attributed to multiple factors. First, direct manipulation interfaces benefit users’ experience by not requiring people to shift their attention from the visual features of interest when interacting~\cite{kondo2014dimpvis, perin:2014:direct}. Second, they simplify the interface by obviating the need for additional control panels or widgets and opens the screen real estate~\cite{beaudouin:2000:instrumental, lee2012beyond,kondo2014dimpvis}.


To help designers create effective tools that implement this form of interaction, there is a need for empirical evidence to guide their design decisions. 
What type of direct manipulation strategies do people perform to convey their interest in performing visualization operations? 
Are there strategies that are consistently employed to perform a given operation?
Are there some that conflict across operations?
Addressing these questions through empirical studies will lead to design principles and guidelines to support further development of visualizations.

We take an initial step to address these questions by conducting a qualitative study in which 10 participants each performed 15 operations on three standard visualizations (scatterplot, bar chart, and histogram).
By using a think-aloud protocol and video analysis, we obtain rich qualitative data 
from which we
extract 
a list of strategies people employ to perform each operation. 
We then identify strategies that have consensus, and strategies that are conflicting with each other. 
Our analysis of the results further sheds light on four high-level categories of strategies (\textit{exemplification}, \textit{declaration}, \textit{instrumentation} and \textit{selection}) from which we derive implications to help designers leverage direct manipulation of graphical encoding. 

This work contributes the following to the area of interaction design in visualization: 
(1) a qualitative characterization of user-defined direct manipulation strategies for performing different operations on visualizations, 
(2) insight into users' mental models when using direct manipulation of graphical encodings as a method for user interaction,
and (3) a set of actionable implications for designing interactive data visualization tools. 
Our results will help designers leverage direct manipulation of graphical encodings as a method for user interaction.

\section{Background}\label{sec:background}
We first discuss the concepts of direct manipulation~\cite{shneiderman:19931} and instrumental interaction~\cite{beaudouin:2000:instrumental}, two key concepts on which research about direct manipulation in visualization is strongly relying. We then discuss the work related to direct manipulation of graphical encodings.

\subsection{Direct Manipulation and Instrumental Interaction}
\textit{Direct manipulation} enables users to directly act on the visual objects of interest~\cite{shneiderman:19931} and provides immediate visual feedback in response to physical actions (e.g., dragging a folder on the desktop using the mouse).
Direct manipulation actions are simple and support continuous flow of interaction. 
\textit{Instrumental interaction}~\cite{beaudouin:2000:instrumental} 
generalizes the principles of direct manipulation by introducing instruments, 
that act as mediators between users and objects of interest. 
For example, sliders or check boxes can be used as instruments for filtering data points. 
Researchers who have applied instrumental interaction principles to visualizations~\cite{lee2012beyond,kondo2014dimpvis,perin:2014:direct,elmqvist2011fluid}; 
highlighted the need to minimize the spatial indirection (the distance between the interaction source and the target object), and the temporal indirection (the delay between invoking an operation and observing the result of that operation)~\cite{perin:2014:direct}.


			



\subsection{Direct Manipulation of Graphical Encodings}
Building on the concepts of direct manipulation and instrumental interaction, \textit{embedded interactions}~\cite{saket:2017:evaluating} are interactions in which users directly manipulate visual marks (e.g., bars in a bar chart) in visual representations rather than widgets and menus to perform a task. The goal of embedded interaction is to tighten the gap between one's intent and the execution of that intent -- thus reduce what Hutchins et al. call the \textit{gulf of execution}~\cite{hutchins1985direct}. It is propitious to \textit{congruent interactions}~\cite{perin:2014:direct}: interactions whose action, reaction, and feedback are cognitively congruent to the intent of the user. These approaches can make interactions more discoverable, resulting in easier to learn and use applications~\cite{perin:2014:direct}. 

There is a growing interest in the infovis and visual analytics communities to develop this form of interaction for visualizations. A large set of visualization authoring tools enable graphic designers to construct customized visualizations (e.g.,~\cite{Liu:2018, Mendez:2016,kim2017data,ren2014ivisdesigner}). These tools use direct manipulation as a method for creating customized visual glyphs. For instance, Data Illustrator~\cite{Liu:2018} enables users to adjust the size of the circle by directly dragging its outline. There are also visualization tools that enable users to perform their analytical tasks using direct manipulation of graphical encodings, for example by letting users convey their interest in sorting a bar chart by dragging the tallest bar to the extreme left or right~\cite{saketVbD,Maulsby:1993,saketVbDTheory}. 
We found 16 papers that explicitly investigate direct manipulation of graphical encodings as a method to let users perform their analytical tasks~\cite{saketVbD, chevalier:2012:histomages, EMS, kondo2014dimpvis, Perin:table, brown2012dis, Perin:2014, Kwon:2011, Saket_VbDEval, saketVbDTheory, riche2010understanding, Wenskovitch2018, desJardins:2007, kim:2016:interaxis, endert2012semantic, torres:2003:visual,Siirtola_Matrix,Baudel:2006}.

We organize this body of work according to which \textit{visualization type} they support. 
We further structure our discussion of each visualization type according to the \textit{encodings} that previous work has explored and the \textit{meanings} that were assigned to their direct manipulation.

\subsection{Visualization type: 2D Scatterplot}
Direct manipulation of position~\cite{saketVbD,kondo2014dimpvis,Baudel:2006}, size~\cite{saketVbD,Kwon:2011}, and color~\cite{saketVbD} of data points in 2D scatterplots have been explored.

\ourParskip
\inlinebold{Position}
Direct manipulation of position in scatterplots has been implemented through drag and drop of data points. 
This has been used to allow users to convey their interest in: navigating the temporal dimension of a dataset~\cite{kondo2014dimpvis}; changing the axes of the scatterplot or to switch to a bar chart~\cite{saketVbD}; and adjusting the values of data points~\cite{Baudel:2006}. 
Researchers in visual analytics have explored direct manipulation of position in scatterplots that represent outputs of statistical models, 
for example to steer clustering models~\cite{desJardins:2007}; 
to steer distance and similarity functions by moving data points closer or further to each other~\cite{brown2012dis, torres:2003:visual, Wenskovitch2018, endert2012semantic}; 
and to define and modify axes of scatterplots by dragging and dropping data points to either side of the x or y axes~\cite{kim:2016:interaxis}.

\ourParskip
\inlinebold{Size}
Direct manipulation of size has been implemented through drag of the outline of a data point's representation. Prior work enabled users to directly manipulate the size of data points shown in a scatterplot to demonstrate their interest in changing the size of data points~\cite{Kwon:2011,saketVbD}, and mapping a data attribute to the size of data points~\cite{saketVbD}.

\ourParskip
\inlinebold{Color}
The direct manipulation of color has been implemented through using a relatively indirect color picker in a contextual menu close to the visual mark. This has been used to change the color of all data points in a scatterplot, and to assign a data attribute to the color of points~\cite{saketVbD}.

\subsection{Visualization type: Bar Chart and Histogram}

Direct manipulation of position~\cite{EMS,saketVbD,Maulsby:1993}, height~\cite{Baudel:2006, kondo2014dimpvis}, and width~\cite{EMS,chevalier:2012:histomages} of bars in bar charts and histograms have been explored.

\ourParskip
\inlinebold{Position}
Direct manipulation of position in these visualizations has been implemented through drag and drop of bars. This has been used to execute abstract operations: to sort a bar chart by dragging the tallest/shortest bar to its extreme left or right side~\cite{saketVbD,Maulsby:1993}; and to merge two bars by dragging and dropping a bar on top of another one~\cite{EMS}. 

\ourParskip
\inlinebold{Height}
Direct manipulation of height has been implemented through drag of the top border of a bar. 
Direct manipulation of the height of bars has been used as a method to convey an interest in navigating the temporal dimension of a dataset~\cite{kondo2014dimpvis}; to adjust the values of data points in bar charts~\cite{Baudel:2006}; and to adjust the weight assigned to an attribute as a method for steering the underlying model used for computing the visual representation~\cite{kim:2016:interaxis}.

\ourParskip
\inlinebold{Width}
Direct manipulation of bar width has been implemented through drag of the left/right borders of a bar to interactively merge and split the bins of a histogram~\cite{EMS}. 

\ourParskip
\inlinebold{Color}
Direct manipulation of color has been implemented with a color picker in a contextual menu close to the visual mark, to change the color of all bars or to assign a data attribute the color of bars~\cite{saketVbD}.





\subsection{Other Visualization Types}


Most research on direct manipulation of graphical encodings has focused on 2D scatterplots and bar charts / histograms, probably due to the simplicity and widespread use of these visualizations.
However, some researchers have also enabled direct manipulation of graphical encodings in other visualization types. This includes the direct manipulation of: \textbf{position} of cells in \textbf{table} visualizations to either steer the underlying ranking model~\cite{Wall_podium} or explore rankings~\cite{Perin:table,Vuillemot:2015}; 
\textbf{angle} of a \textbf{pie chart} segment to navigate the time dimension~\cite{kondo2014dimpvis};
rows and columns in \textbf{matrix} visualizations~\cite{Siirtola_Matrix,perin:2014:bertifier}; 
nodes in \textbf{tree} visualizations~\cite{vuillemot:2016:tournaments}; 
and \textbf{position} of tokens in \textbf{unit} based visualizations~\cite{huron2014constructing}.

\begin{table*}[ht]
\footnotesize
\caption{The 15 basic operations that have been used in previous work for direct manipulation of graphical encodings in 2D scatterplot, bar chart and histogram. 
We use all 15 operations in our study. 
The last column (\textbf{Phrasing}) contains the exact sentence participants were told in our study. 
All the operations started with ``\textit{How would you interact with this system to show that you are interested in:} ''}
\begin{center}
\begin{tabular}{p{0.06\textwidth}p{0.08\textwidth}p{0.34\textwidth}p{0.42\textwidth}}
     \textbf{Encoding} & \textbf{Visualization} & \textbf{Operation} & \textbf{Phrasing} \\ 
    \hline
    \multirow{6}{*}{Position} 
    & \multirow{4}{*}{Scatterplot} 
    & \command{1} Assign a data attribute to an axis~\cite{saketVbD, kim:2016:interaxis}  \newline 
      \command{2} Switch from a scatterplot to a bar chart~\cite{saketVbD} \newline
      \command{3} Navigate the values of a point over time~\cite{kondo2014dimpvis}\newline
      \command{4} Adjust the value of a point~\cite{Baudel:2006}
    & 
      Assigning the horsepower attribute to the x-axis \newline
      Switching from a scatterplot to a bar chart \newline
      Checking if this specific car has ever had the price of 20,000?\newline
      Adjusting the value of this specific car to 30,000
      \\
     \arrayrulecolor{gray}
     \cline{2-4}
     \arrayrulecolor{black}
    & \multirow{2}{*}{Bar chart} 
    & \command{5} Group the bars into one bar~\cite{EMS} \newline 
      \command{6} Sort the bar chart~\cite{saketVbD,Maulsby:1993} 
    & Merging the bars representing SUV and Wagon cars\newline
      Sorting the bar chart in an ascending order
      \\
    \hline 
    \multirow{2}{*}{Size} 
    & \multirow{2}{*}{Scatterplot} 
    & \command{7} Change the size of all points~\cite{saketVbD,Kwon:2011}  \newline 
      \command{8} Assign a data attribute to the size of points~\cite{saketVbD} 
    & Change the size of all data points, so that they are all equally bigger  \newline
      Assigning city mile per gallon attribute to the size of points
         \\
    \hline
    
    \multirow{5}{*}{Color} 
    & \multirow{2}{*}{Scatterplot}
    & \command{9} Change the color of all points ~\cite{saketVbD}  \newline 
      \command{10} Assign a data attribute to the color of all points~\cite{saketVbD} 
    & Changing the color of all points to red \newline 
      Assigning the cylinder attribute to the color of all points.
     \\
     \arrayrulecolor{gray}
     \cline{2-4}
     \arrayrulecolor{black}
    & 
    \multirow{2}{*}{Bar chart} 
    & \command{11} Change the color of all bars ~\cite{saketVbD}  \newline 
      \command{12} Assign a data attribute to the color of all bars~\cite{saketVbD} 
    & Changing the color of all bars to red \newline 
      Assigning the car type attribute to the color encoding
         \\
    \hline 
    \multirow{2}{*}{Height} 
    & \multirow{2}{*}{Bar chart}
    &  \command{13} Navigate the values of a point over time~\cite{kondo2014dimpvis}\newline
     \command{14} Adjust value of a bar~\cite{Baudel:2006}
    & Checking if the number of sedan cars have ever had the value of 35?
  \newline
      Adjusting the number of the SUV cars to 15
      \\
    \hline 
    \multirow{1}{*}{Width} 
    & \multirow{1}{*}{Histogram}
    & \command{15} Expand the range of a bin in a histogram~\cite{EMS}  
    & Expanding the range of this specific bin from 2009 to 2010?
    \\
    \hline
\end{tabular}
\end{center}
\label{tab:commands}
\end{table*}

\subsection{The Next Step: Collecting Empirical Data}

Because direct manipulation of graphical encodings is a recent topic of research, existing work has explored sparse points in the design space by selecting \textit{operations} to be invoked and decided which \textit{strategies} should be used to invoke each operation. It is now appropriate to empirically study these designs. Table~\ref{tab:commands} summarizes the operations and associated strategies from previous work for 2D scatterplots, bar charts, and histograms. Although we focus on visualization operations and discard analytical ones, this list gives a starting point for gathering empirical data regarding how people accomplish such operations.


\section{Preliminary Studies}\label{sec:pilot}

We conducted two preliminary studies to determine which approach to use to empirically investigate direct manipulation of encodings.

\subsection{Preliminary Study 1: Paper-Based Study}
In our first pilot study, we asked three participants (2 male, 1 female) to verbally explain how they would perform a series of visualization tasks using direct manipulation of graphical encodings on paper prints of visualizations. 
Providing participants with paper-based visualizations makes it possible to remove constraints that come with any implemented system, thus give more freedom and expressivity to participants.

We printed a bar chart representing the Cars dataset~\cite{henderson1981building}. 
We explained the concepts of marks, encodings, labels, and direct manipulation of graphical encodings to the participants. We then explained the visualization and the data. 
We gave each participant three operations to perform in a random order \textit{(e.g., \textit{How would you show that you are interested in sorting the bar chart in an ascending order?})}. 
We asked participants to verbalize how they would perform each operation only using direct manipulation of the encodings used in the visualization.

Overall, participants found it challenging to explain their strategies 
without being able to \textit{actually} perform the operations. 
For example, one participant said: \textit{``Should I imagine that I can change the width of the bar? [...] then what is the system response?''}, and another \textit{``I have no idea what happens if I move this bar!''} 
Moreover, participants often did not restrict themselves to direct manipulation of graphical encodings only. 
For example, to sort the bar chart, one participant said he would first drag the tallest bar to the right side of the bar chart and added: \textit{``Now I expect to see a drop-down menu that has the `sort' option.''} -- thus combining direct manipulation of graphical encodings and WIMP.

Based on the results of this first study, we decided to provide an interactive tool that supports direct manipulation of different graphical encodings used in visual representations. 
This would ideally help us avoid participants' confusion as to what interactions are available to them and how those interactions are implemented. In addition, we could encourage the participants to think of strategies that rely solely on direct manipulation of encodings, if no other interactions are available.

\subsection{Preliminary Study 2: Partial Implementation}

We conducted a second pilot study with four new participants (4 male) using a prototype with limited functionality.
We developed a web-based interactive bar chart showing the Cars dataset~\cite{henderson1981building} and supporting  direct manipulation of graphical encodings used in bars (e.g., changing the height of a bar). 
We explained the visualization, data, and available interactions to the participants. 
Then we asked them to perform the same three operations as in the first pilot study using the system. 

Participants were able to perform 8 out of 12 operations (4 participants $\times$ 3 operations) using direct manipulation of graphical encodings. Although participants found some operations challenging to invoke (e.g., assigning a new data attribute to the axis), this pilot study indicated that using an interactive prototype is an appropriate way of eliciting people's strategies for performing operations using direct manipulation of graphical encodings.

\section{Study Design}\label{sec:study}
In this study, we investigate \textit{how} people perform the visualization operations listed in Table~\ref{tab:commands} using direct manipulation of graphical encodings. We describe the visualization and encoding types we used, the operations from Table~\ref{tab:commands}, the dataset we used in the study and the software implementation. Then we describe our participants and settings, study procedure, and data collection method and analysis.

\subsection{Visualization and Encoding Types}


Given its qualitative and observational nature, 
we included a small number of (three) visualization types in the study to keep it within reasonable time.
We selected \textbf{scatterplot}, \textbf{bar chart} and \textbf{histogram} because:
i) they are among the most commonly used visualizations~\cite{few2009now}; and 
ii) direct manipulation of graphical encodings has been well researched with these visualizations, as shown in Table~\ref{tab:commands}.
We studied the five encodings that have been explored in previous work for these visualizations: \textbf{position} and \textbf{color} for scatterplot and bar chart; \textbf{size} for scatterplot; \textbf{height} for bar chart; and \textbf{width} for histogram.


\subsection{Operations}

We included in our study all 15 operations, \command{1} to \command{15}, listed in Table~\ref{tab:commands}. 
We made this choice to compare the strategies implemented in previous work for each operation with the strategies our participants would perform. 
We also made the choice to focus on operations for visualization construction (i.e., that are designed to specify the visualization), and excluded operations that are designed to steer underlying models used for computing visualizations (e.g., InterAxis~\cite{kim:2016:interaxis}).


\begin{figure*}[t]
\centering
    \includegraphics[width=\linewidth]{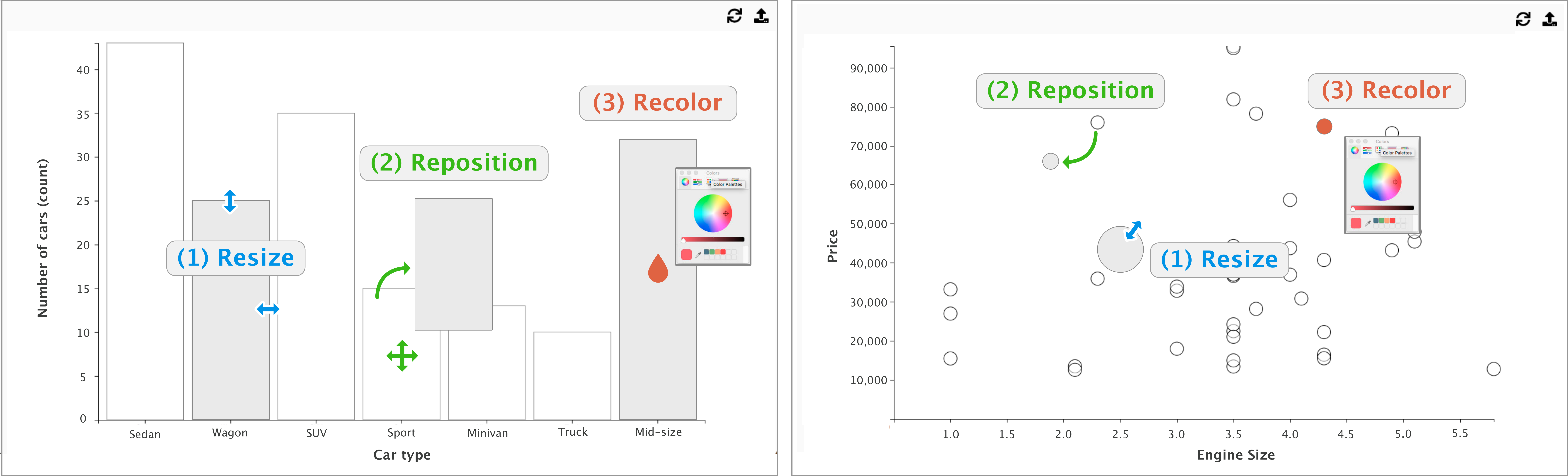}
  \caption{This figure shows our study platform and supported interactions for different visualizations: bar chart (left) and scatterplot (right). Users can \textcolor{resize}{\textbf{(1) resize}}, \textcolor{reposition}{\textbf{(2) reposition}}, and \textcolor{recolor}{\textbf{(3) recolor}} bars and points directly.
  }
  \label{fig:interactions} 
\end{figure*} 

\subsection{Datasets}
Most operations listed in Table~\ref{tab:commands} are generic enough to not depend on the dataset being used. 
For example, assigning a data attribute to an axis can be performed on any scatterplot regardless of the underlying tabular dataset. 
Operations that involve navigating temporally, however, require a dataset that contains at least one temporal dimension.
Based on this constraint, we selected the widely used cars~\cite{henderson1981building} and movies~\cite{TableauData} datasets, which have also been used in related studies (e.g.,~\cite{saketVbD,kim:2016:interaxis,kondo2014dimpvis}).

\subsection{Interactive Tool and Interactions}
We developed an interactive visualization tool using JavaScript and the D3 library~\cite{d3:infovis11}. 
An Upload (\faUpload) button supports uploading the pre-defined visualizations we used in our study.
A Reset button (\faRefresh) supports resetting the visualization.
We added interactivity to the encodings used in visualizations. 
For instance, participants could change the position, color, width, and height of the bars in a bar chart. 
Or, they could change the size, position, and color of circles in a scatterplot. 

The tool shows one visualization at a time. 
Participants can directly manipulate the encodings in the visualization. 
For example, after the interviewer has uploaded a bar chart and asked a participant to perform an operation, the participant can use any of the provided interactions to convey their intention to perform this operation.
The tool only enables participants to manipulate the encodings and does not recompute the visualization (similar to a drawing interface like Adobe Illustrator). 
Enabling the participants to convey their actions without the system reacting to those actions makes it possible to observe participants' unrevised behavior, and drive system design to accommodate it.

To add interactivity to the five graphical encodings (position, height, width, size, color) investigated in this study, we kept our implementation of interactions as close as possible to previous work (Figure \ref{fig:interactions} illustrates these interactions for the bar chart and the scatterplot):


\ourParskip
\begin{enumerate}[noitemsep,nolistsep]
    \item \inlinebold{Position} We afforded the repositioning of circular data points in a scatterplot or bars in a bar chart or histogram anywhere on the screen through drag and drop.
    \item \inlinebold{Height and Width} Clicking on a bar in a bar chart or histogram makes four small handles (black circles) appear on the four sides of the bar. Dragging the left and right handles changes the width of the bar; dragging the top and bottom handles changes its height.
    \item \inlinebold{Size} Clicking on a circular data point in a scatterplot makes a small handle (black circle) appear on the perimeter of the circle. Dragging the handle changes the size (radius) of the circle.
    \item \inlinebold{Color} Right clicking on a mark (circular data point in scatterplot, bar in bar chart or histogram) makes a color picker appear. Picking a color from the color picker changes the color of the mark.
\end{enumerate}




\subsection{Participants and Settings}

We recruited 10 non-color blind participants (4 females, 6 males), aged 20--30 (mean $24.8$) via email and word of mouth at our university.
They were 
students enrolled in computer science (5), physics (1), psychology (2), and mechanical engineering (2). 
All reported being familiar with reading visualizations, and eight had created visualizations before. 
Four took the information visualization course taught at our university. 
Some participants had experience with visualization tools such as Microsoft Excel (8), D3.js (4) and Processing (1).
None of them had participated in the pilot studies.
Participants sat approximately 30--40 cm from a 13'' LCD display with a resolution of 2560 $\times$ 1600 pixels equipped with a mouse and keyboard. The visualizations were shown in full screen. 

Given the qualitative nature of the study, we determined participant numbers based on empirical saturation~\cite{creswell1998qualitative} -- which can be reached with as low as 6 participants~\cite{morse2000determining}. 
Two authors of this paper watched the screen-recorded videos after each session to get a sense of the strategies the participant used to perform the operations. In addition, during each session the interviewer took notes of high-level strategies that the participant employed to perform each operation. As we progressed through our study, we discussed these notes, identifying whether the observed strategies were repeats or newly observed strategies. The sessions with participants 9 and 10 generated limited new strategies, suggesting that we had reached empirical saturation. We then discussed our informal findings as a group and decided to conclude the study.



\subsection{Procedure}

\inlinebold{1. Introduction (\texttildelow 10 min)}
Participants were briefed about the purpose of the study and their rights.
After filling out the study consent form and a questionnaire on demographics and visualization expertise, 
they watched a three-minute video explaining the concepts of marks, encodings, labels, and axes. 
Participants could replay the video as they liked. 
Then, they were given a sample of the movies dataset printed on a sheet of paper. 
After the experimenter had explained to them the meanings of rows and columns, 
they were asked to familiarize themselves with the data for two minutes and ask any question they might have.

\ourParskip
\inlinebold{2. Training (\texttildelow 10 min)}
In this phase, participants were shown a scatterplot, a bar chart, and a histogram all created with the movies dataset (one visualization at a time, in a random order). 
We explained how the system supports manipulation of different encodings used in each visualization. 
For instance, we showed participants that they can drag the left or right boundary of a bar to manipulate its width. We then asked participants to perform an operation on the visualization. 
Depending on the visualization, we asked participants to: 
assign a data attribute to an axis of a scatterplot; 
sort a bar chart in a descending order; 
and expand the range of a bin in a histogram. 
For example, the interviewer asked the participants: \textit{``How would you interact with this system to show that you are interested in sorting the bar chart in a descending order?''} 
The interviewer's role in this phase was not to guide the participant, but solely to answer their questions. 
Such questions included: \textit{``Am I supposed to show how I am going to do this task by manipulating these glyphs''? ``Should I assume that the system is going to detect my interaction?''} 
The interviewer did not suggest participants strategies for performing the given operations, nor gave examples or hints on how to perform an operation.


\ourParskip
\inlinebold{3. Main study (\texttildelow 30 min)}
Participants familiarized themselves with the cars dataset like they had with the movies dataset. 
Then, they were shown a scatterplot, a bar chart, and a histogram created with the cars dataset, one at a time with the order randomized. 
For each visualization, they were asked to perform all operations associated with the visualization (see Table~\ref{tab:commands}), one at a time in a random order.
In total, each participant performed 15 operations ({\footnotesize{ \textit 8 with the scatterplot \textbf{+}  6 with the bar chart \textbf{+} 1 with the histogram}}).
For example, the interviewer asked the participants: 
\textit{``How would you interact with this system to show that you are interested in changing the color of all points to red?''} 
Participants were asked to perform each operation by only manipulating the graphical encodings in the visualization on the screen. Participants could also verbally explain how they would perform the operation when they could not perform it with the supported interactions. They could also suggest more than one strategy for performing each operation.

\ourParskip
\inlinebold{4. Wrap-up (\texttildelow 5 min)}
The experimenter thanked the participants who received a \textdollar 10 gift card.
Participants were invited to ask additional questions about the study.

\begin{figure*}[!ht]

\begin{minipage}[t]{\linewidth}
\centering
\includegraphics[width=1.0\linewidth]{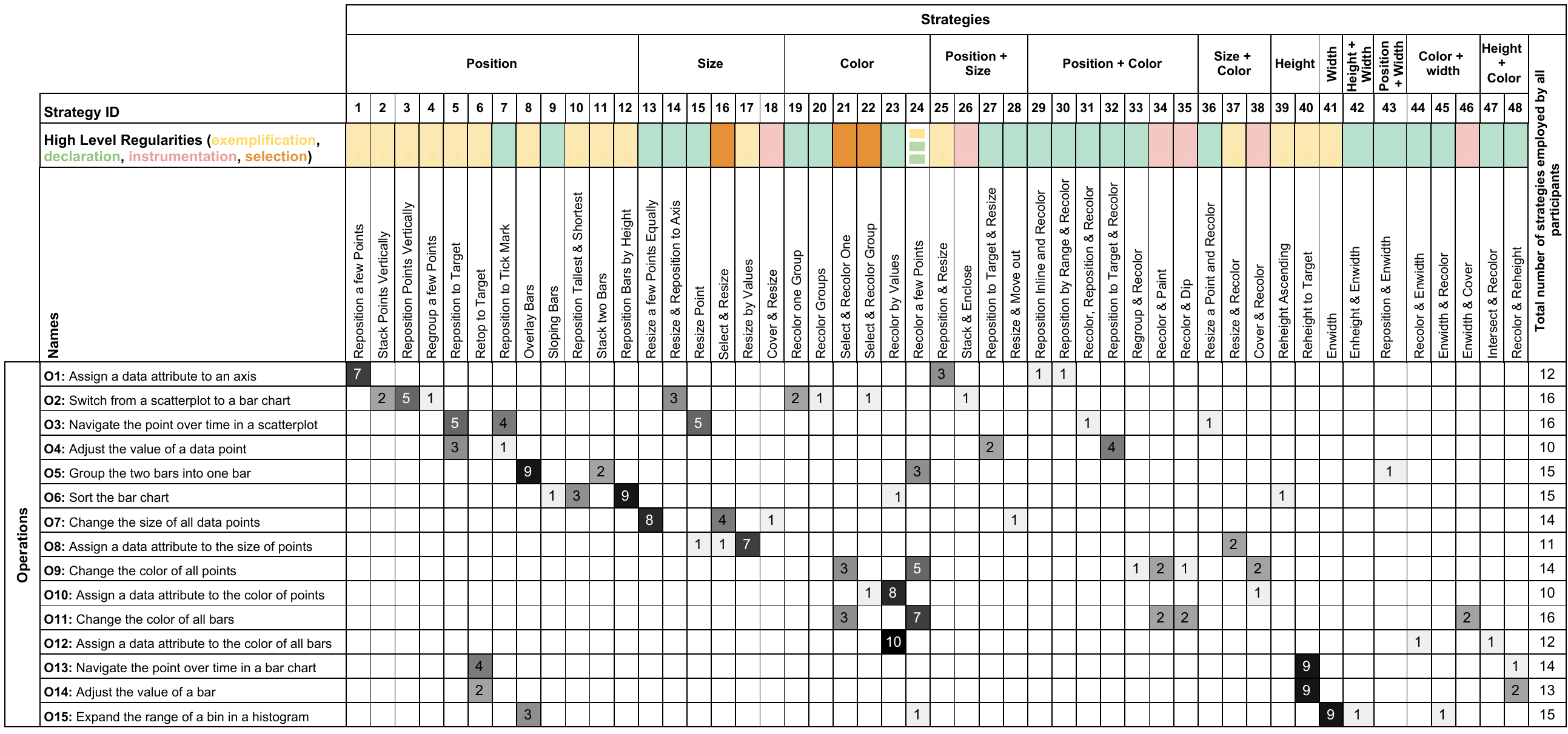}
\vspace{-1em}
  \caption{
    Each row is one of the 15 operations participants performed during the study, and each column is one of the 48 strategies we identified. 
    Each cell shows the number of times participants used the strategy in column to perform the operation in row. The higher the value in a cell, the darker the background of the cell. 
    Strategies are grouped based on the main encoding(s) involved in employing that strategy (second row in the table). 
    For each strategy we color code the high-level approaches: exemplification, declaration, instrumentation and selection (fourth row in the table), detailed in the Discussion section. We provide detailed description of each strategy in Figure~\ref{fig:strategiesExplanation}. 
    }
    \vspace{1.5em}
  \label{fig:meaning}
\end{minipage}
\begin{minipage}[t]{\linewidth}
\centering
\includegraphics[width=1\linewidth]{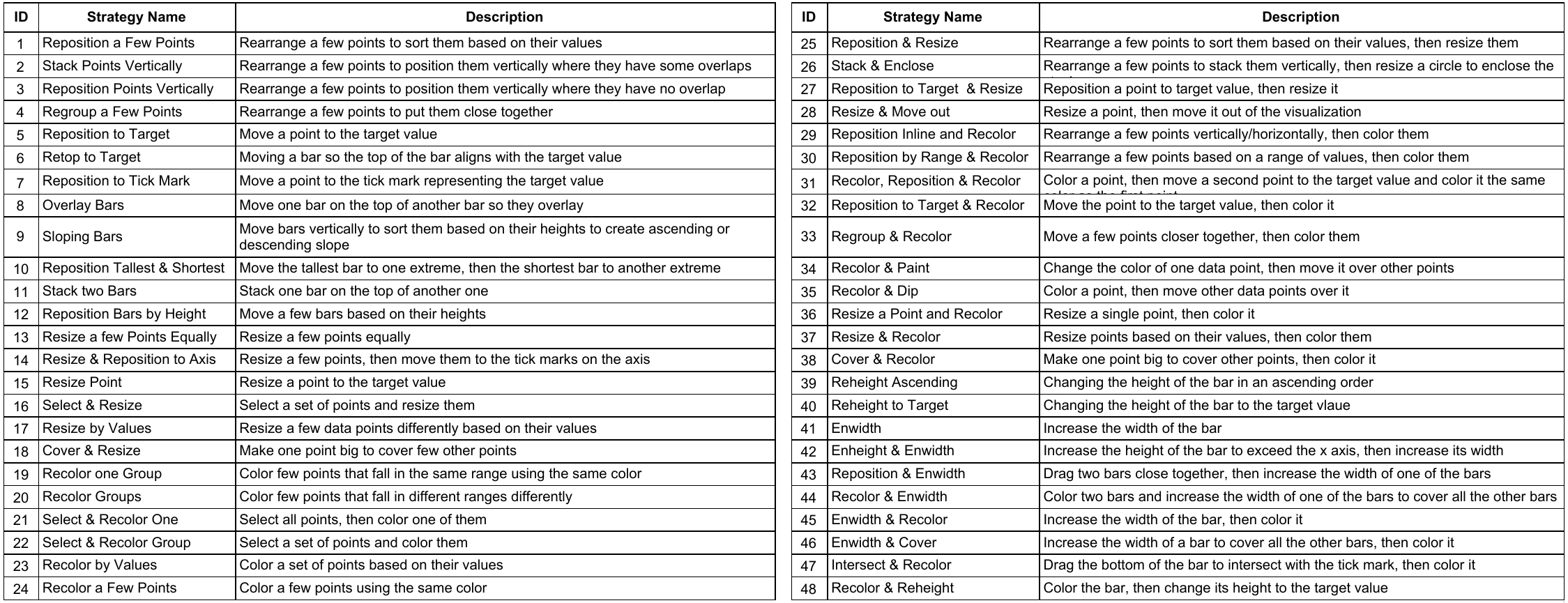}
\vspace{-1em}
  \caption{
    Description of each of the 48 strategies participants used in our study.}
    \vspace{0.5em}
  \label{fig:strategiesExplanation}
\end{minipage}

\end{figure*}

\subsection{Data collection and Method of Analysis}
We screen- and audio-recorded the study.
During the main study, the experimenter took notes of participants' actions.
We analyzed the 298 minutes of screen-capture videos using open coding~\cite{creswell2002educational} in three phases. 

In the \textit{familiarization} phase, two coders reviewed the videos together to identify common and unexpected patterns that informed our decision to focus the analysis on strategies. For example, they observed that participants came up with many more unique ways of switching the visualization from a scatterplot to a bar chart, than for other operations.

In the \textit{coding} phase, once the two coders had a good understanding of the data, they agreed that strategies for performing operations were an appropriate lens through which to analyze the data and defined the `intended strategies' as the unit of analysis. This meant that two types of information had to be coded for: which operation the participant is performing; and which intended strategy the participant either employs or suggests. An intended strategy is the expression of an intention that a participant performed physically and/or explained verbally to complete an operation. This definition allowed us to capture intended strategies that participants demonstrated using the prototype as well as those that they verbally explained, e.g., \textit{``In that case, I would select the entire screen [all data points]. Then once they are in the same group of selection, I am assuming that increasing the size of one object should increase the size of all.''} To ensure a reliable coding process, both coders coded parts of a video together until they fully agreed on the coding. Then one coder went through all videos and coded both operations and intended strategies. The second coder then checked the coding for two randomly selected videos and confirmed the coding. In total 203  \textit{intended strategies} were coded across all 15 operations.

In the \textit{analysis} phase, both coders watched the video snippets of the 203 intended strategies to derive \textit{archetypal strategies}. 
For each instance of intended strategy, they discussed whether it fell under any of the existing archetypal strategies or if it formed a new one. 
For every new archetypal strategy, the coders sketched a storyboard of it on paper. 
This process created 48 mutually exclusive archetypal strategies. 
The two coders further named each archetypal strategy based on its sketch. 
Figure~\ref{fig:meaning} and Figure~\ref{fig:strategiesExplanation} summarize these 48 archetypal strategies (that we call `strategies’ in the remainder of the paper).

We provide online
\footnote{\url{https://encodingstudy.github.io/}} all relevant materials for this study: datasets, software for running the experiment, and sketches of the strategies employed by the participants.

\section{Results}\label{sec:results}
We present the strategies participants employed in our study to perform the operations using direct manipulation of graphical encodings. All but one participants could identify at least one strategy for each operation, with one participant failing to identify a strategy for one operation only. 
Participants suggested 4--5 strategies per operation on average.
Most strategies are based on direct manipulation of the graphical encodings using well-known interaction patterns such as \textit{repositioning}, \textit{resizing} (that includes resizing the radius of points, the width of bars and the height of bars), and \textit{recoloring}. 
A few strategies, however, rely on direct manipulation of other visualization elements such as axes, labels, and tick marks (participants verbally explained these strategies).


We identified 48 unique strategies across the 15 operations, shown in Figure~\ref{fig:meaning} (the exact phrasing used in our study for each operation i provided in Table~\ref{tab:commands}).
Below, for each operation we present the commonalities, variety, and unexpectedness of the strategies that participants employed.  We provide raw sketches drawn during the video analysis procedure in supplemental materials. For simplicity, in this section we use ``points'' to refer both to the data cases in the dataset and to the visual marks representing these data points in the scatterplot (circles in our study); we use ``bars'' to refer to the visual marks representing data points or aggregations of data points in the bar chart and the histogram.

\begin{figure}[!t]

\begin{minipage}[t]{1.0\linewidth}
	\centering
	\includegraphics[width=1.0\linewidth]{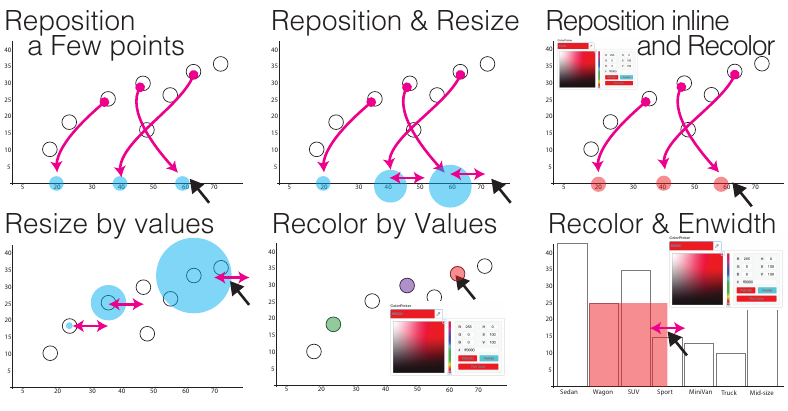}
	\vspace{-1.0\baselineskip}
	\caption{Prominent strategies to assign a data attribute to an axis (first row); and to the size or color of points in the scatterplot, and of bars in the bar chart (second row).}
	\label{fig:attributingStrategies}
\end{minipage}

\vspace{2.5\baselineskip}

\begin{minipage}[t]{1.0\linewidth}
	\centering
	\includegraphics[width=1.0\linewidth]{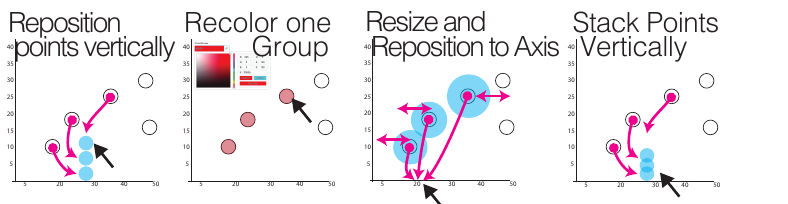}
	\vspace{-1.0\baselineskip}
	\caption{Strategies to switch from scatterplot to bar chart.}
	\label{fig:switchscatterplotbarchart}
\end{minipage}

\vspace{2.5\baselineskip}

\begin{minipage}[t]{1.0\linewidth}
	\centering
	\includegraphics[width=1.0\linewidth]{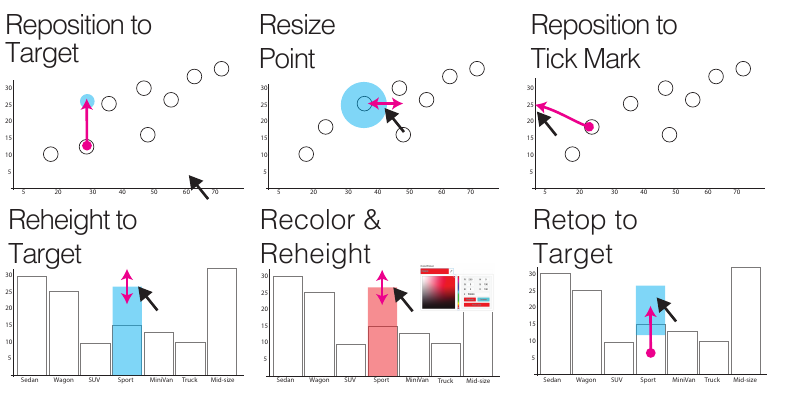}
	\vspace{-1.0\baselineskip}
	\caption{Strategies to navigate a data point over time.}
	\label{fig:navigatetime}
\end{minipage}

\vspace{2.5\baselineskip}

\begin{minipage}[t]{1.0\linewidth}
	\centering
	\includegraphics[width=1.0\linewidth]{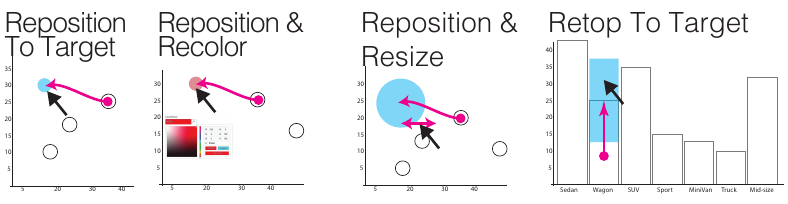}
	\vspace{-1.0\baselineskip}
	\caption{Four strategies to adjust the value of a point in a scatterplot and of a bar in a bar chart.}
	\label{fig:adjustingStrategies} \vspace{-1em}
\end{minipage}

\vspace{1.5\baselineskip}

\end{figure}

\inlinebold{Assign a data attribute to an axis (\commandHeader{1}), to the size of points (\commandHeader{8}), and to the color of points (\commandHeader{10}) or bars (\commandHeader{12})}
We grouped together these operations that are about assigning a data attribute to an encoding. 
Figure~\ref{fig:attributingStrategies} shows the main strategies used for invoking these operations.


We identified four strategies that participants employed to assign a data attribute to a scatterplot axis (\command{1}). 
Seven participants used \strategy{Reposition a few Points}, where they rearranged a few points to sort them based on their values. 
This strategy has already been supported in previous work~\cite{saketVbD}.
Three participants used \strategy{Reposition \& Resize}, where they first rearranged a few points to sort them horizontally and then resized them based on their values. 
Here they manipulated the size of points to express their intent to sort the points from lower to higher values. For example, after sorting and resizing the points, P2 said: \textit{``So the size represents the horsepower, but at the same time the position is also horsepower. By the size [changing the size] I am trying to reinforce that the higher value should be on the right."} 


One participant used \strategy{Reposition inline \& Recolor}, where he repositioned a few data points to sort them horizontally and then colored the points based on their values to inform the system about his thought process. He stated: \textit{``I am coloring them just to show the system they have been sorted.''} Another participant also first repositioned a few data points to sort them based on the range that they fall in and then colored them (\strategy{Reposition by range \& recolor}). 

To assign a data attribute to the size of points in the scatterplot (\command{8}), 
seven participants used \strategy{Resize by Values}, where they resized a few points based on their values. 
For example, P2 made smaller three points with a low value; then he made the point with the highest value bigger than all the other.
Other participants used similar strategies: P1 resized a single random point (\strategy{Resize point}), and P2 and P8 first resized a few points based on their values before coloring them (\strategy{Resize \& Recolor}). 
One participant also said that he would first select all the points and then resize one of the selected points (\strategy{Select \& resize}).

Most participants employed \strategy{Recolor by Values} to assign a data attribute to color, both with the scatterplot (\command{10}, 8 participants) and with the bar chart (\command{12}, 10 participants). This strategy consists of coloring a few points/bars with different values using different colors. 
P5 suggested \strategy{Recolor \& Enwidth} with the bar chart, where he first colored two bars then increased the width of one of the bars to cover all the remaining bars. He explained: \textit{``if I want to imply changes to be made to all bars, I drag the width of the bar.''}



\ourParskip
\inlinebold{Switch from a scatterplot to a bar chart (\commandHeader{2})}
Figure~\ref{fig:switchscatterplotbarchart} shows the four most used strategies to invoke this operation out of the eight we identified. 
Five participants used \strategy{Reposition Points Vertically}, where they rearranged a few points to position them vertically without any overlap. 
Two used \strategy{Recolor one Group}, where they colored a few points that fall in a specific range using the same color. 


Three used \strategy{Resize \& Reposition to Axis}, where they resized a few points then moved them to the tick marks shown on the axis. 
They explained that they consider each resized point to be a bar where the value is mapped to the radius of the circle instead of the height of the bar, e.g., \textit{``this looks like the bar chart the bars are circular'' (P5)} and \textit{``So basically each circle is a bar'' (P2).} 
Two participants used \strategy{Stack Points Vertically}, where they rearranged a few points to position them vertically where they have some overlaps (i.e., stacking the points vertically similar to visualization by demonstration~\cite{saketVbD}).


\ourParskip
\inlinebold{Navigate a data point over time (\commandHeader{3} \& \commandHeader{13})}
Figure~\ref{fig:navigatetime} shows the main strategies used to check whether a data point has ever had a target value.
With the scatterplot (\command{3}),
five participants used \strategy{Reposition to Target}, where they moved the point to the target value like in DimpVis~\cite{kondo2014dimpvis}; 
five used \strategy{Resize Point}, where they resized the data point; 
and four used \strategy{Reposition to Tick Mark}, where they moved the point and dropped it on the tick mark representing the target value. 

With the bar chart (\command{13}),
nine used \strategy{Reheight to Target}, where they changed the height of the bar to the target value like in DimpVis~\cite{kondo2014dimpvis};
one used \strategy{Recolor \& Reheight}, where she first changed the height of the bar to the target value and then colored the bar;
and four used \strategy{Retop to Target}, where they moved the bar vertically so that the top of the bar aligns with the target value. 




\ourParskip
\inlinebold{Adjust the value of a point (\commandHeader{4}) or a bar (\commandHeader{14})}
Figure~\ref{fig:adjustingStrategies} shows the most frequent strategies for these operations. 
Participants used four different strategies to adjust the value of a point in the scatterplot to a target value (\command{4}). 
Three used \strategy{Reposition to Target}, where they moved the point to the target value, like in previous work~\cite{Baudel:2006}. 
Six participants extended this strategy by adding another step to it: four further colored the point (\strategy{Reposition \& Recolor}) and two other participants resized it (\strategy{Reposition \& Resize}).


To adjust the value of a bar in the bar chart (\command{14}), nine participants changed the height of the bar to the target value (\strategy{Reheight to Target}), like in previous work~\cite{Baudel:2006}. 
Two used \strategy{Retop to Target}, where they moved the bar so that the top of the bar aligns with the target value.

\begin{figure}[!t]

\begin{minipage}[t]{1.0\linewidth}
	\centering
	\includegraphics[width=1.0\linewidth]{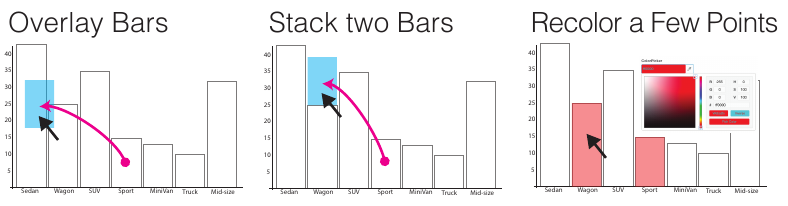}
	\vspace{-1.0\baselineskip}
	\caption{Three strategies to group two bars into one bar.}
	\label{fig:Mergetwobars}
\end{minipage}

\vspace{2.0\baselineskip}

\begin{minipage}[t]{1.0\linewidth}
	\centering
	\includegraphics[width=1.0\linewidth]{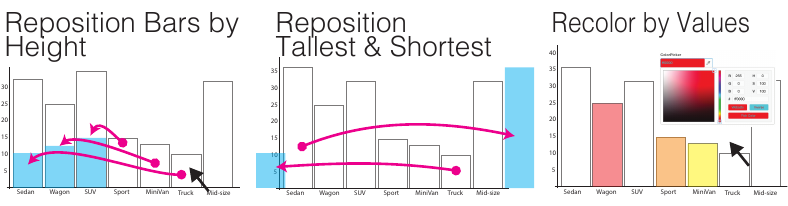}
	\vspace{-1.0\baselineskip}
	\caption{Three strategies to sort a bar chart.}
	\label{fig:sortingStrategies}
\end{minipage}

\vspace{2.0\baselineskip}

\begin{minipage}[t]{1.0\linewidth}
	\centering
	\includegraphics[width=1.0\linewidth]{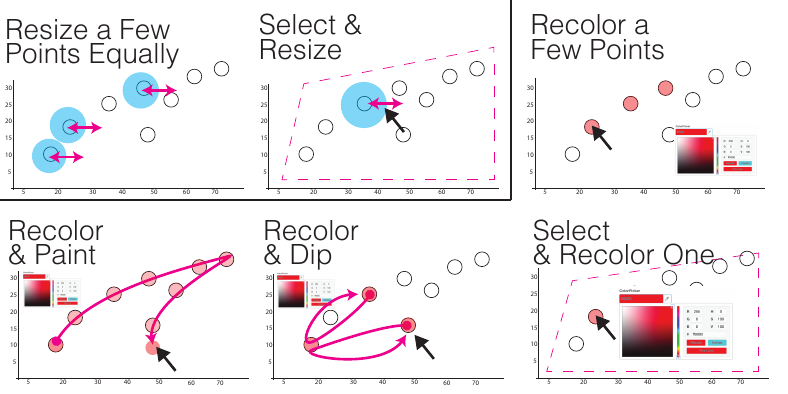}
	\vspace{-1.0\baselineskip}
	\caption{Strategies to change the size or color of all points in a scatterplot, and the color of all bars in a barchart.}
	\label{fig:resizeRecolorStrategies}
\end{minipage}

\vspace{2.0\baselineskip}

\begin{minipage}[t]{1.0\linewidth}
	\centering
	\includegraphics[width=1.0\linewidth]{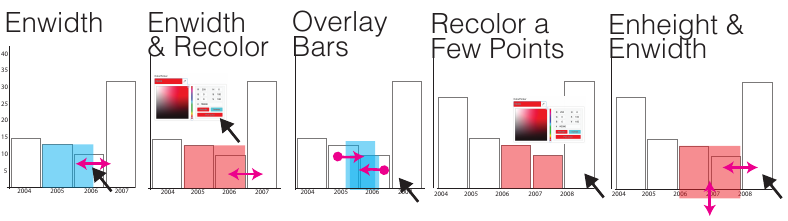}
	\vspace{-1.0\baselineskip}
	\caption{Strategies to expand the range of a bin.}
	\label{fig:binstrategy}
\end{minipage}

\end{figure}

\ourParskip
\inlinebold{Group two bars into one bar (\commandHeader{5})}
Figure~\ref{fig:Mergetwobars} shows the three strategies (out of four) used by more than one participant. 
Nine used \strategy{Overlay Bars}, where they dragged one bar on the other one, like this has been proposed in previous work~\cite{EMS}. 
Participants found this strategy to be intuitive, e.g., \textit{``Is there an easier way to this question? There might be others but this seems like the most intuitive one'' (P1).} 
Two used \strategy{Stack two Bars}, where they stacked one bar on top of another, e.g., \textit{``after grouping two bars, values of the two bars should add up'' (P3).} Here we use ``stack'' to indicate the piling of visual marks on top of each other without overlap.
Three participants used \strategy{Recolor a few Points}, where they colored the two bars using the same color. 



\ourParskip
\inlinebold{Sort a bar chart (\commandHeader{6})}
Figure~\ref{fig:sortingStrategies} shows three of the five strategies participants suggested to sort a bar chart in ascending order. 
Nine participants used \strategy{Reposition Bars by Height}, where they moved a few bars based on their heights. For example, P1 stated: \textit{``dragging two or three bars one after another should essentially indicate that I am rearranging them based on their values.''} 
Three participants used \strategy{Reposition Tallest \& Shortest}, where they dragged the tallest bar to one extreme of the visualization and the shortest bar to another extreme, like in previous work~\cite{saketVbD,Maulsby:1993}. 
Some of them felt that moving both tallest and shortest bars was not necessary. For instance, P3 said: \textit{``I pick sedan which is the biggest one and put it on here [extreme right]. This makes sense and then the system should be able to pick up what I was trying to do. And maybe if you want to make it more sure, you pick up the lowest one also put it here [extreme left].''} P6 used \strategy{Recolor by Values}, where she colored three different bars with ordinal colors. She explained: \textit{``keep the first one white, then make the second one yellow. So I am assuming this is the increasing order'' (P6).}




\ourParskip
\inlinebold{Change the size of all points (\commandHeader{7}) or the color of all points (\commandHeader{9}) / bars (\commandHeader{11})}
Figure~\ref{fig:resizeRecolorStrategies} shows the main strategies participants used when asked to execute these operations.

There were two popular strategies to change the size of all circles in the scatterplot (\command{7}).
Eight participants used \strategy{Resize a few Points Equally}, where they resized two to four random points to make them equally bigger. 
For example, P2 stated: \textit{``If I have to change the size of all of them [data points in scatterplot], then probably I should make a group of circles and they should not share the same axis, they should be very random.''} 
Four suggested verbally \strategy{Select \& Resize}, that consists of selecting all circles and resizing one of them.


There were four main strategies to change the color of all points in the scatterplot (\command{9}) or bars in the barchart (\command{11}).
Many participants used \strategy{Recolor a Few Points}, where they recolored two to three points/bars using the same color (five did so with the scatterplot, seven with the bar chart). 
P1 and P6 used \strategy{Recolor \& Paint}, where they colored one point/bar and dragged it over a few other points/bars. 
Similarly, P5 used \strategy{Recolor \& Dip}, where he first colored one point/bar, then dragged a few other points/bars such that they overlaid the colored point/bar. 
Finally, six participants mentioned that if selection was available they would use \strategy{Select \& Recolor One}
, where they would first select all points/bars then color one of them. 


\ourParskip
\inlinebold{Expand the range of a bin in a histogram (\commandHeader{15})}
Participants suggested five strategies to expand the range of a bin from one range to another range. Figure~\ref{fig:binstrategy} shows three of these strategies.


Nine participants used \strategy{Enwidth}, where they increased the width of a bar to merge it with the bars next to it, as has been proposed in previous work~\cite{EMS}. 
P7 extended this strategy into \strategy{Enwidth \& Recolor}, where she first increased the width of the bar and then colored the bar yellow to inform the system about the latest updates: \textit{``[...] I colored the bar to show that it has been updated'' (P7).} 
Three participants used \strategy{Overlay bars}, where they dragged and dropped the bins (bars) falling in the expected target range (e.g., the bars between 2008-2011) on top of each other such that they overlaid. 
P2 used \strategy{Recolor a few points}, where he colored in the same color the bins that fell in the expected target range.
P3 used \strategy{Enheight \& Enwidth}, where he first increased the height of the bar to exceed the $x$ axis then increased its width to merge the bar with the bars next to it.




\section{Discussion}\label{sec:discussion} 

We structure our discussion based on Figure~\ref{fig:meaning}, that presents the 48 strategies (in columns) participants used to perform the 15 operations \command{1} to \command{15} (in rows). 
We first discuss the varying degrees of agreement for strategies as well as conflicting strategies (strategies that participants used to perform different operations). 
Then, we propose four high-level approaches for organizing the 48 strategies: \textit{exemplification}, \textit{declaration}, \textit{instrumentation} and \textit{selection}.
Last, we discuss limitations of our study and future research directions.


\subsection{Varying Degrees of Agreement}
Some strategies were widely adopted by participants for a given operation, i.e. they have a relatively high degree of agreement. 
For two thirds of the operations (10/15), one strategy represents more than 50\% of the distribution of all used strategies. 
For example, among the 12 strategies that were proposed to assign a data attribute to the color of all bars, 10 employed \strategy{Recolor by Values} (see \command{12}, strategy 23). 
As another example, among the 15 strategies participants proposed to expand the range of a bin in a histogram, 9 employed \strategy{Enwidth} (see \command{15}, strategy 41). 
On the other hand, some operations were performed using a wide range of strategies, with no consensus emerging. 
For example, to switch from a scatterplot to a bar chart (O2),
participants used 8 different strategies. This suggests that it might be
harder to identify common strategies for certain operations.

Findings from our empirical study show that users of visualizations can effectively employ direct manipulation of graphical encoding as a means of performing operations of varying complexity. However, the degree of agreement regarding which strategy to use varies from operation to operation. Going forward, visualization designers might consider incorporating  strategies in consensus for performing operations using direct manipulation of graphical encodings. As such, our study provides a framework for collecting more empirical data and contributes to building a set of operations and corresponding consensual strategies.

\subsection{Strategies in Conflict}
Participants sometime employed the same strategy to perform different operations. For example, participants colored a few visual marks using the same color (\strategy{Recolor A Few Points}) to perform a variety of operations including: 
to \textit{group two bars in one bar} (\command{5}, strategy 24); 
to \textit{change the color of all points} in both the scatterplot (\command{9}, strategy 24) and the bar chart (\command{11}, strategy 24); 
and to \textit{expand the range of a bin in a histogram} (\command{15}, strategy 24).


The fact that participants used the same strategy to perform different operations, in a relatively open environment, is revealing in several ways.
First, it suggests that the space of strategies for performing operations using direct manipulation of encodings is not as vast as it might look.
Second, it indicates that there is often no single strategy that will be unanimously used to perform an operation. 
Third, it shows that some strategies can be consensual for performing different operations. 
That is the case for \strategy{Reheight to Target}, for example, which was used 9 times both to navigate the point over time in a bar chart (\command{13}, strategy 40) and to adjust the value of a bar (\command{14}, strategy 40).

This many to many relationship between strategies and operations raises technical challenges in leveraging direct manipulation of graphical encoding for user interaction. Visualization designers should in the first stage avoid implementing support for strategies that different people would use for performing different operations. Future work should explore recommending to the user all possible operations in response to an employed strategy so that the user can select the most appropriate operation (similar to VisExemplar~\cite{saketVbD}). 
Not only would this enable designers to support multiple operations, this could also be used to collect data on people's preferences towards developing an understanding of contextual strategies. 
In other words, collecting such data would enable us to analyze in which situations a given strategy is intended to trigger a particular operation.

\subsection{Higher level Categorization of Strategies}
By comparing the relations between strategies and operations, we identified four high-level approaches that participants used to manipulate encodings to invoke an operation (see Figure~\ref{fig:meaning}). Below we discuss these four high-level approaches: \textit{exemplification}, \textit{declaration}, \textit{instrumentation} and \textit{selection}.

\ourParskip
\inlinebold{Exemplification} With this approach, participants directly manipulated a small number of visual marks in order to \textit{illustrate by example} the output they were trying to achieve. Participants widely used \textit{exemplification} to invoke operations (e.g., see Rows 1-6 in Figure~\ref{fig:meaning}). Exemplification was mostly achieved through a repetitive set of actions to show the system how a part of the visual output should look like. For instance, one of the strategies that participants used to sort the bar chart is \strategy{Reposition Bars by Height} (strategy 12). With this strategy, they positioned a few bars one by one in an ascending order similar to how the bar chart should look like after sorting. In another example, one of the strategies that participants used to change the color of all data points is \strategy{Recolor a Few Points} (strategy 24). With this strategy, participants colored only a subset of points using the same color to demonstrate their higher-level goal of changing the color of all data points. The idea behind \textit{exemplification} is similar to the visualization by \textit{demonstration} paradigm~\cite{saketVbD}. In visualization by demonstration, one of the methods to provide visual demonstrations is to directly manipulate the graphical encodings used in the visualization to indicate a part of the expected visual output to the system. 

\ourParskip
\inlinebold{Declaration} 
With this approach, participants manipulated a graphical encoding different from the primary graphical encoding used in the operation. For example, when we asked to assign a data attribute to an axis of a scatterplot (\command{10}), we expected participants to mainly rely on positioning data points manually based on their values for that data attribute, since switching the axis results in changes in how points are positioned. 
However some participants manipulated encodings that were not directly linked to assigning a data attribute to the axis. 
One participant used \strategy{Reposition Inline and Recolor} (strategy 29), where they first rearranged a few points vertically/horizontally, then colored them. 
While repositioning a few points is directly related to mapping a data attribute to \textit{position}, coloring the points is not directly related to creating such a mapping. 
In this case, participants used color as a way to communicate their intention to the system. 
Another participant used \strategy{Enwidth \& Recolor} to expand the range of a bin in a histogram (\command{15}). While color is not directly related to how ranges in histograms are represented, the participant colored the bar as a way to inform the system about the latest changes they had made.
Another participant used \strategy{Recolor by Values} to sort the bar chart. 
Here, while the primary graphical encodings related to the sort operation are the height and the position of bars, the participant expressed the notion of order using an ordinal color scheme.

\ourParskip
\inlinebold{Instrumentation} 
With this approach, participants used a visual mark as an instrument (or tool). 
This resulted in relatively advanced and unexpected strategies.
For example, two participants used \strategy{Recolor and Paint} to change the color of all points (\command{9}) and bars (\command{11}).
They first colored a point/bar and then used this colored visual mark as a brush: they dragged it over other points/bars to color them.
Similarly, two other participants used \strategy{Recolor and Dip} to perform these two same operations (\command{9} and \command{11}), but the other way around.
They first colored a point/bar and then used this colored visual mark as a bucket: they dragged a few other points/bars such that they overlaid the colored point/bar.
To turn a visual mark into an instrument, participants used the graphical encodings \textit{color} (strategies 34 and 35) and \textit{size} (strategies 18, 38 and 46).
The idea behind instrumentation is similar to what has been proposed with constructible interfaces~\cite{tools}, a paradigm that is strongly grounded into instrumental interaction~\cite{beaudouin:2000:instrumental}. 
In constructible interfaces, one of the methods to create or modify visual marks is to turn other visual marks into instruments that can be used to perform operations~\cite{tools}.

\ourParskip
\inlinebold{Selection} 
With this approach, participants expressed their interest in a selection technique (verbally because selection was not supported in the prototype). 
They explained that they would prefer having access to a selection option that they could use to select a subset of points. 
Then they would apply an operation to a single data point rather than having to apply the same operation to multiple points.
For example, participants used \strategy{Select \& Recolor One} to change the color of all points in both the scatterplot and the bar chart.

\ourParskip
\inlinebold{Differences in approaches in terms of visual marks} 
We can further differentiate these four approaches according to the number of visual marks they require one to manipulate. Participants sometimes thought of strategies that relied on direct manipulation of \textit{three or more} graphical encodings. In such cases, participants thought of \textit{instrumentation} and \textit{selection} approaches because these approaches enabled them to perform repetitive actions rapidly (instead of manipulating different visual marks one by one). 
For example, when asked to change the size of all data points, some participants mentioned that resizing every single point might be tedious. Thus, they suggested \strategy{Select \& Resize} where they would select a subset of points and resize them all at once rather than resizing each point individually.
On the other hand, participants used \textit{instrumentation} and \textit{selection} less often when their strategies required them to change one or two visual marks. It suggests that the number of marks to directly manipulate to perform an operation informs which high-level approach to use.

\subsection{Types of Encodings Used for Performing an Operation}
We initially hypothesized that participants would manipulate multiple encodings while employing their strategies. For example, we thought participants would manipulate a variety of encodings such as color, size, and height of the bars while applying their strategies to sort a bar chart. However, our results show that a majority of strategies often rely on  manipulating a single type of graphical encoding. For instance, for 13 out of the 15 strategies employed to sort a bar chart, participants only manipulated the \textit{position} of the bars. Similarly, 13 out of the 14 strategies employed to map a data attribute to the size encoding, participants only manipulated the \textit{size} of the data points. 

This indicates that users of visualizations can effectively employ direct manipulation of graphical encoding as a means of performing operations, and that the mechanisms people might use to manipulate graphical encodings is not as complex and wide as one could expect. 
These findings can also help tool designers decide whether to implement direct manipulation of only one, or of multiple encodings in a tool according to the operations the tool must support.


\subsection{Limitations}
This qualitative study is the first attempt to understand how people manipulate graphical encodings to invoke different visualization operations. 
As such, it cannot answer all open questions related to this problem; here we discuss the limitations of our study and findings.

\ourParskip
\inlinebold{Prototype Functionality} 
Our findings must be interpreted in the context of the study prototype. 
To investigate how people directly manipulate graphical encodings to perform operations, we had to first design a tool that supports such functionality. 
Many participants explained strategies that were not supported by the prototype. 
This shows that participants thought about strategies beyond the implemented interactions, as we had hypothesized.
However, it is likely that the limited functionality of our study software has impacted their strategies. Building on our findings, future studies should consider including some of the functionality suggested by our participants, such as the ability to select data points. 
This will allow participants to express more diverse strategies and broaden the palette of possible ways of performing visualization operations through direct manipulation of graphical encodings.

\ourParskip
\inlinebold{Operation Description} 
Although we kept the description of the operations as close as possible to previous work, the phrasing of the questions can influence the strategies employed by participants. Moreover, the nature of the operations differ from one another. 
For example, some operations are more direct in how they refer to the changes that need to be made on the visual marks (i.e., the questions are congruent~\cite{Boy_Literacy} to the task to achieve). 
Despite this limitation, we observed a multiplicity of strategies for all operations.

\subsection{Generalizability and Future Work} 
Our qualitative study provides rich and detailed results; however this study does not aim at generalization. 
Testing our research questions using other visualization techniques and operations might reveal nuances of direct manipulation of graphical encodings that were not tested in this study. 
One of the next steps will be to build on our findings to study how people perform and perceive a large set of operations and strategies via a crowdsourced study in order to gather more quantitative data and aim for more generalizable results. 

In this study, we investigated how people \textbf{perform} different operations using direct manipulation of encodings. We aim to extend our study by investigating how people \textbf{understand} this method for user interaction. For example, we plan to conduct a study where we show videos of user interactions from this study to Amazon Mechanical Turk workers and ask them to list potential operations they think the person in the video is trying to perform. We expect that after collecting such data, we will be able to design machine learning models that predict people's intended operations based on the strategies they employ.

The prototype used in our study only enabled participants to manipulate the visual marks and their graphical encodings -- it did not react to participants’ actions (similar to a drawing interface like Adobe Illustrator). 
Enabling participants to convey their actions without the system reacting to those actions enabled us to observe participants’ unrevised behavior in isolation. 
However, a setup in which the system would react to participants' actions for performing an operation might influence strategies employed by the participants. 
Specifically, if the system responds to each individual action then accurate predictions of operations may shorten strategies employed by the participants. 
We envision conducting a Wizard of Oz experiment in which we will provide different types of feedback as participants manipulate the encodings to perform an operation. This will help us understand how the reaction of the system influences the strategies employed by participants.

Today, visualization tools incorporate multimodal interactions to enhance user experience and system usability~\cite{badam2017affordances}. 
For instance, earlier work on natural language interfaces for visualization indicated potential value in combining direct manipulation and natural language as complementary interaction techniques~\cite{Setlur:2016}. Exploring how direct manipulation of graphical encodings combines and complement other input modalities such as natural language, body posture, and gestural interaction is an exciting avenue for future research.

Interactivity in many visualizations has been, and is, supported through WIMP widgets. These tools are successful in easing the processes of visualization construction, because they allow users to interactively construct visualizations instead of using programming and are fully discoverable. However, when other forms of interaction such as direct manipulation of graphical encodings are leveraged, it raises a number of important questions~\cite{saket2019liger} including: \textit{How effective are different forms of interaction (e.g., WIMP vs. direct manipulation) for specific operations? How can we design tools that effectively leverage several forms of interaction?} 
Understanding the differences and trade-offs between various forms of interactions and how they are used for specific operations will help designers and developers make informed interaction design decisions when creating visualization tools.


\section{Conclusions}
Designing visualizations that leverage the direct manipulation of graphical encodings as a means for user interaction is challenging because we lack a holistic understanding of which actions should be supported and what are their associated operations. To date, such systems rely on designer-generated heuristics to create these mappings. This paper provides the first list of strategies, sometimes consensual and sometimes conflicting, that people employ to perform operations using direct manipulation of graphical encodings. We organized the strategies into four high-level approaches: \textit{exemplification}, \textit{declaration}, \textit{instrumentation}, and \textit{selection}. This work provides a framework for collecting more empirical data and contribute towards developing an holistic understanding of the strategies people employ to perform certain operations.

\acknowledgments{
We are grateful to the study participants, reviewers, and members of the Georgia Tech Visualization Lab for their helpful insights and comments. This project is supported by NSF IIS-1750474 and the Natural Sciences and Engineering Research Council
of Canada (NSERC).}

\bibliographystyle{abbrv-doi}

\bibliography{template}
\end{document}